\input harvmac
%\noblackbox
%\input epsf

%%% Figures

%\newcount\figno
%\figno=0
%\def\fig#1#2#3{
%\par\begingroup\parindent=0pt\leftskip=1cm\rightskip=1cm\parindent=0pt
%\baselineskip=11pt

%\global\advance\figno by 1
%\midinsert
%\epsfxsize=#3
%\centerline{\epsfbox{#2}}
%\vskip 12pt
%\centerline{\vbox{{\bf Figure \the\figno:} #1}}\par
%\endinsert\endgroup\par}
%\def\figlabel#1{\xdef#1{\the\figno}}
%\overfullrule=0pt

%%% macros

\def\underarrow#1{\vbox{\ialign{##\crcr$\hfil\displaystyle
 
{#1}\hfil$\crcr\noalign{\kern1pt\nointerlineskip}$\longrightarrow$\crcr}}}
% use of underarrow
%A~~~\underarrow{a}~~~B
%
\def\tilde{\widetilde}

%
%%% Paragraphs

%%% special math symbols
\font\cmss=cmss10
\font\cmsss=cmss10 at 7pt
\def\rlx{\relax\leavevmode}
\def\inbar{\vrule height1.5ex width.4pt depth0pt}
\def\IC{\relax\,\hbox{$\inbar\kern-.3em{\rm C}$}}
\def\IN{\relax{\rm I\kern-.18em N}}
\def\IP{\relax{\rm I\kern-.18em P}}
\def\IR{\relax{\rm I\kern-.18em R}}
\def\IC{{\relax\hbox{$\inbar\kern-.3em{\rm C}$}}}
\def\IZ{\relax\ifmmode\mathchoice
{\hbox{\cmss Z\kern-.4em Z}}{\hbox{\cmss Z\kern-.4em Z}}
{\lower.9pt\hbox{\cmsss Z\kern-.4em Z}}
{\lower1.2pt\hbox{\cmsss Z\kern-.4em Z}}\else{\cmss Z\kern-.4em
Z}\fi}
\def\IH{\relax{\rm I\kern-.18em H}}
\def\ZZ{\rlx\leavevmode\ifmmode\mathchoice{\hbox{\cmss Z\kern-.4em Z}}
 {\hbox{\cmss Z\kern-.4em Z}}{\lower.9pt\hbox{\cmsss Z\kern-.36em Z}}
 {\lower1.2pt\hbox{\cmsss Z\kern-.36em Z}}\else{\cmss Z\kern-.4em
 Z}\fi}
%%% misc.
\def\narrowplus{\kern -.04truein + \kern -.03truein}
\def\narrowminus{- \kern -.04truein}
\def\narrowminussub{\kern -.02truein - \kern -.01truein}

\def\frac#1#2{{#1\over #2}}

\def\IZ{\relax\ifmmode\mathchoice
{\hbox{\cmss Z\kern-.4em Z}}{\hbox{\cmss Z\kern-.4em Z}}
{\lower.9pt\hbox{\cmsss Z\kern-.4em Z}}
{\lower1.2pt\hbox{\cmsss Z\kern-.4em Z}}\else{\cmss Z\kern-.4em
Z}\fi}

%
%       \eqn\label{a+b=c}	gives displayed equation, numbered
%				consecutively within sections.
%     \eqnn and \eqna define labels in advance (of eqalign?)
%
\def\eqnn#1{\xdef #1{(\secsym\the\meqno)}\writedef{#1\leftbracket#1}%
\global\advance\meqno by1\wrlabeL#1}
\def\eqna#1{\xdef #1##1{\hbox{$(\secsym\the\meqno##1)$}}
\writedef{#1\numbersign1\leftbracket#1{\numbersign1}}%
\global\advance\meqno by1\wrlabeL{#1$\{\}$}}
\def\eqn#1#2{\xdef #1{(\secsym\the\meqno)}\writedef{#1\leftbracket#1}%
\global\advance\meqno by1$$#2\eqno#1\eqlabeL#1$$}

%%%%%%%%%%%%%%%%%%%%
%%%%%%%%%%%%%%%%%%%%

%%%%%%%%%%%%%%%%%%%%

\def\mnk{{\cal M}_{N,k}}

\def\tX{{\tilde X}}

\nref\cds{A. Connes, M.R. Douglas and A. Schwarz,
``Noncommutative Geometry and Matrix Theory: Compactification on Tori''
hep-th/9711162, JHEP 9802 (1998) 003}

\nref\cnsbook{A. Connes, ``Noncommutative Geometry'', Academic Press 1994}

\nref\lennydip{D. Bigatti and L. Susskind,
``Magnetic fields, branes and noncommutative geometry'',
hep-th/9908056.}

\nref\sw{N. Seiberg and E. Witten,
``String Theory and Noncommutative Geometry'',
hep-th/9908142, JHEP 9909 (1999) 032.}

\nref\zy{Z. Yin, 
``A Note on Space Noncommutativity'',
hep-th/9908152, Phys. Lett. B466, 234-238, 1999.}

\nref\mh{H. Liu and J. Michelson, 
``Stretched Strings in Noncommutative Field Theory'',
hep-th/0004013,  Phys. Rev. D62 (2000) 066003}

\nref\lnynti{N. Seiberg, L. Susskind and N. Toumbas,
``Strings in Background Electric Field, Space/ Time Noncommutativity and 
A New Noncritical String Theory'',
hep-th/0005040, JHEP 0006 (2000) 021}

\nref\hahrvr{R. Gopakumar, J. Maldacena, S. Minwalla and A. Strominger,
``S-Duality and Noncommutative Gauge Theory'',
hep-th/0005048, JHEP 0006 (2000) 036}

\nref\eliezer{J.L.F. Barbon and E. Rabinovici, 
``Stringy Fuzziness as the Custodian of Time-Space Noncommutativity'',
hep-th/0005073, Phys. Lett. B486, 202-211, 2000.}

\nref\orisav{O.J. Ganor, G. Rajesh and S. Sethi,
``Duality and Non-Commutative Gauge Theory'',
hep-th/0005046}

\nref\om{R. Gopakumar, S. Minwalla, N. Seiberg and A. Strominger,
``OM Theory in Diverse Dimensions''
,hep-th/0006062, JHEP 0008 (2000) 008}

\nref\brga{E. Bergshoeff, D. S. Berman, J. P. van der Schaar and P. Sundell,
``Critical fields on the M5-brane and noncommutative open strings'',
hep-th/006112}

\nref\brgb{E. Bergshoeff, D. S. Berman, J. P. van der Schaar and P. Sundell,
``A Noncommutative M-Theory Five-brane'', hep-th/0005026}

\nref\myers{R.C. Myers,
``Dielectric Branes'', hep-th/9910053, JHEP 9912 (1999) 022}

\nref\giant{J. McGreevy, L. Susskind and N. Toumbas,
``Invasion of the Giant Gravitons from Anti-de Sitter Space'',
hep-th/0003075}

\nref\agm{O. Aharony, J. Gomis and T. Mehen,
``On Theories With Light-Like Noncommutativity'',
hep-th/0006236.}

\nref\stromem{A. Strominger, 
``Open P-Branes'',
hep-th/9512059, Phys. Lett. B383 (1996) 44}

\nref\abs{O. Aharony, M. Berkooz and N. Seiberg,
``Light-Cone Description of (2,0) Superconformal Theories in Six Dimensions'',
hep-th/9712117, Adv. Theor. Math. Phys. 2 (1998) 119}

\nref\bfss{T. Banks, W. Fischler, S.H. Shenker and L. Susskind,
``M Theory As A Matrix Model: A Conjecture'',
hep-th/9610043, Phys. Rev. D55 (1997), 5112}

\nref\lnydlcq{L. Susskind,
``Another Conjecture about M(atrix) Theory'',
hep-th/9704080}

\nref\abkss{O. Aharony, M. Berkooz, S. Kachru, N. Seiberg and E. Silverstein,
``Matrix Description of Interacting Theories in Six Dimensions'',
hep-th/9707079, Adv. Theor. Math. Phys 1 (1998) 148}

\nref\eshmon{O. Aharony, M. Berkooz, S. Kachru and E. Silverstein,
``Matrix Description of (1,0) Theories in Six Dimensions'',
hep-th/9709118, Phys. Lett. B420, (1998) 55;
S. Kachru, Y. Oz and Z. Yin,
``Matrix Description of Intersecting M5-branes'',
heo-th/9803050, JHEP 9811 (1998) 004}

\nref\orijo{O.J. Ganor and J.L. Karzmarek,
``M(atrix)-Theory Scattering in the Noncommutative (2,0) theory'',
hep-th/0007166}

\nref\clmem{B. de Wit and H. Nicolai, 
``Supermembarnes: A Fond Farewell?'',
Contribution to Conf. on Supermembranes and Physics in 2+1-Dimensions, 
Trieste, Italy, Jul 17-21, 1989;
B. de Wit, U. Marquard and H. Nicolai,
``Area Preserving Diffeomorphisms and Supermembrane Lorentz Invariance'',
Commun. Math. Phys.128:39, 1990; 
B. de Wit, M. Luscher and H. Nicolai,
``The Supermembrane is Unstable'',
Nucl.Phys.B320:135,1989;
B. de wit, J. Hoppe and H. Nicolai,
``On the Quantum Mechanics of Supermembranes'',
Nucl. Phys. B305:545,1988} 

\nref\base{T. Banks, N. Seiberg and S. Shenker,
``Branes from Matrices'',
hep-th/9612157, Nucl. Phys. B490 (1997) 91.}

\nref\opnlc{ K. Ezawa, Y. Matsuo and K. Murakami,
``Matrix Regularization of an Open Supermembrane ---towards M-theory
five-branes via open supermembranes'',
hep-th/9707200, Phys. Rev. D57 (1998) 5118}

\nref\mdfour{M. Berkooz,
``From SYM Perturbation Theory to Closed Strings in Matrix Theory'',
hep-th/9912241,JHEP 0005 (2000) 050}

\nref\nikita{H.W. Braden and N.A. Nekrasov,
``Space-Time Foam From Non-Commutative Instantons'', 
hep-th/9912019}

\nref\newk{K. Lee, D. Tong and S. Yi,
``The Moduli Space of Two U(1) Instantons on Noncommutative $R^4$ and 
$R^3\times S^1$'',
hep-th/0008092;
K.-Y. Kim, B.-H. Lee and H.-S. Yang,
``Comments on Instantons on Noncommutative $R^4$'',
hep-th/0003093}

\nref\mttwo{K. Furuuchi,
``Instantons on Noncommutative $R^4$ and Projection Operators'',
hep-th/9912047;
}

\nref\havgrp{M. Spradlin, A. Strominger and A. Volovich, to appear}

%%%%%%%%%%%%%%%%%%%%

\Title{\vbox{\hbox{hep-th/0010158}\hbox{PUPT-1964}\hbox{RUNHETC-2000-40}}
} {\vbox{\centerline{Light-like (2,0) Noncommutativity and}
\centerline{}
\centerline{Light-Cone Rigid Open Membrane Theory}}}
\smallskip
\centerline{Micha Berkooz\footnote{$ $} {mberkooz@feynman.princeton.edu} }
\vskip 0.09in
\medskip\centerline{$ $ \it Joseph Henry Laboratories, 
Princeton University,
Princeton, NJ 08544}
\medskip\centerline{$ $ \it Department of Physics and Astronomy, 
Rutgers University, Piscataway, NJ, 08855}

%\centerline{\it Institute for Advanced Study}\centerline{\it
%Princeton, NJ 08540, USA}

\vskip 0.30in

\centerline{\bf Abstract}

The six-dimensional (2,0) field theory admits a generalized
``noncommutative'' deformation associated with turning on a large null
3-form field strength. This theory is studied using its discrete
light-cone formulation as quantum mechanics on a blow-up of the ADHM
moduli space. We show how to interpret the ADHM manifold as
configurations of open membranes, and check our results against basic
space-time considerations.

%\draftmode
\vskip 0.05in

\Date{September 2000}
\vfill\eject

%%%%%%%%%%%%%%%%%%%%%%%%%%%%%%%%%%%%%%%%%%%%%%%%%%%%%%%%%%%%%%%%%%%%%%%%

\newsec{Introduction}

Quantum theories on non-commutative spaces \cds\cnsbook\ are typically
theories of extended objects \lennydip\sw\zy. The familiar case is that of a
2-form turned on in a gauge theory, in which case the relevant objects
are 1+1 dimensional - either rigid dipoles in the case of space-space
non-commutativity
\lennydip\mh\ or fluctuating strings  in 
the time-space case \lnynti\hahrvr\eliezer (see also \orisav). 
A similar mechanism, however, operates when
turning on other field strengths, i.e, some higher dimensional
extended object (depending on which field strength is turned on)
becomes light as the field strength becomes large. In the context of
theories decoupled from gravity this was discussed for example in
\om\brga\brgb, and in the context of theories with gravity initially in 
\myers\giant. 

In this note we will discuss the deformation of the six-dimensional
(2,0) field theory by a 3-form field strength. More precisely we will
discuss a deformation by a null 3-form field strength \agm, i.e., a
field of the form
\eqn\dfrm{H_{+ij}\not=0} where $X^+$ is a lightcone coordinate, 
and $i,j$ 
are some transverse space-like coordinates\foot{We will be more
specific about this later.}. In this case one expects that the
extended object that will become light is a open membrane
\om\brga\brgb. If we realize this system as a cluster of M5-branes in
11 dimensional M-theory then the open membrane is the
after-decoupling-remnant of an M2-brane ending on the M5-brane
\stromem. Note that this theory is not OM theory \om\brga\brgb\ because 
the field strength is null and not time like. We will touch on this more 
at the end of the introduction. 

The existence of such a theory has been known for some time now. In
\abs\ a discrete light cone quantization of this theory was suggested,
in the spirit of \bfss\lnydlcq.  This description is a generalization
of the DLCQ description of the (2,0) CFT \abkss\ (or theories with
lower susy
\eshmon).  Indeed this model was recently used \orijo\ to partially
substantiate the extended object nature of the theory by computing
leading corrections to the free action of a single tensor multiplets for 
the case of a single M5-brane.
In this paper we will pursue further this description and show how to
extract from it information about the configurations and fluctuations
of these large membranes. 

This is actually only an approximate statement since in this theory
there is no truely well defined notion of ``the state of a
membrane''. As for the (2,0) field theory, this theory has no
dimensionless couplings which control the strength of the interactions
- all the interactions are of strength one at some energy
scale\foot{For a single 5-brane there is a valid low momentum
expansion}. Hence the configurations of a single connected membrane
mix strongly with configurations in which there are many smaller
membranes. A more precise statement would be to find the wave
functional on the space of open membranes which describes an asymptotic
scattering state. We will do so later.

Finally, let us comment that turning on a null field strength, as
explained in \agm, is not the generic field strength which can be
turned on \sw. In particular it is not OM theory \om\brga\brgb. In the
null theory the particles are replaced by membranes with a small set
of dynamical fluctuations. The membranes will have rotational degrees
of freedom and their boundary perhaps fluctuates in a limited way, but
there are no excitations in the interior of the membrane - i.e., their
bulk is rigid. Hence, overall, the number of degrees of freedom
relative to the (2,0) CFT is not radically increased. This is to be
contrasted with OM theory in which one believes that there is at least
some proceeses in which one will see fully fluctuating open membranes,
including fluctuations of the interior. Specializing to this case will
enable us on the one hand to have better control of the model, and on
the other hand we expect that much of what we say (the open membrane
interpretation of the fields in the discrete light cone quantization)
will be applicable to the general case.

The organization of the paper is the following. In section 2 we review
the set-up and the decoupling limit. We then gauge fix the open
membrane to the light cone and extract some basic information about
its ground state. We will later compare these results to those of the
discrete light cone. In section 3 we review the DLCQ of this
theory following \abs, show how to derive it from discretizing open
membranes, and discuss how to measure the volume of the membrane, its
distribution moments, and its ground state wave function. In
section 4 we discuss some special examples of points on the ADHM
manifold and their open membrane interpretation.

A alternative approach for the quantization of open membranes, and the
corresponding derivation of a Matrix model, will be discussed in
\havgrp.

\newsec{The NCG (2,0) and its Membranes}

\subsec{The Decoupled Theory}

The kinematical set-up was discussed in \abs\agm\ and we will only
review it briefly here. The background fields that we will turn on are
\eqn\bck{\matrix{ G^{\mu\nu}&=\eta^{\mu\nu}, \cr
		  H_{12+}=H_{34+}&\not=0. }} As explained in \agm\
this configuration of $H$ satisfies the self-duality equations for the
3-form field strength on the brane.  Next we need to specify the
decoupling limit. This is simplest to understand in the case of a
single M5-brane\foot{We will refer to the $k$ M5-brane loosely as
``the $U(k)$ case''.}. Before turning on the 3-form field, the
worldvolume action for a single M5-brane is the free action for a
(2,0) tensor multiplet. After turning it on, the action is no longer
free but it still has a long wave length approximation.  As suggested
in
\agm\ the criterion is to keep finite a dimension 9 operator correction 
to the free action (the operator is discussed in \orijo. It is closely
related to the extended object nature of the theory). Hence we need to
keep finite a dimension -3 combination of the background $H$ field and
$M_p$ which will be the coefficient of this operator. This gives us
the decoupling limit \abs\agm
\eqn\dcplim{M_p\rightarrow\infty,\ \ \ \ H/M_p^6=\ finite. }

\subsec{The Membranes}

As mentioned above when one turns on a field strength, what used to be
point like particles blows up into extended objects. To which branes
they blow up depends on the details of the field strength(s) turned
on, but on general grounds these branes are closed and hence only
generate multipole moments (generalizing the non-commutative geometry
dipoles
\lennydip\ which appear when turning on a B field). In our case, the
ordinary point like particles of the (2,0) multiplet are replaced by
open membranes whose boundaries lie on the M5-brane (clearly the
theory has been modified from a free theory to an interacting one
because, even after decoupling, these open membranes now have dipole
charges under the 2-form vector potential on the brane). In this
section we will perform a heuristic effective action analysis of these
membranes which we will later compare to the discrete light cone
predictions.

The light cone formulation of the closed membrane was discussed in
\clmem\ (as well as its regularization by matrices, which yields the 
BFSS matrix theory Lagrangian).
The light cone gauge fixing condition is defined by
\eqn\lcc{X^{\pm}=\sqrt{{1\over2}}(X^0\pm X^1)}
\eqn\lccb{X^+(\tau,\sigma_1,\sigma_2)=X^+(0)+\tau,}
and the light cone action is
\eqn\endlag{w^{-1}{\cal L}=
{1\over 2}(D_0X)^2-{1\over 4}\{X^a,X^b\}^2,
}
where $w$ is a 2D measure normalized to 1. The covariant derivative is 
defined $$D_0X=\partial_0 X^a - \{\alpha,X\},$$ where $\alpha$ is 
a gauge field for area preserving diffeomorphisms, which is a gauge symmetry 
of this Lagrangian: 
\eqn\apgg{\sigma^r\rightarrow\sigma^r+\beta^r(\sigma),\ i=1,2,\ \ \ \ 
\partial_r(w(\sigma)\beta^r(\sigma))=0.}
which we can write locally as 
$$\beta^r(\sigma)={\epsilon^{rs}\over w(\sigma)}\partial_s\beta(\sigma).$$
The transformation rules of the fields are
$$\delta X^a=\{\beta,X^a\},\ \ 
  \delta{\alpha}=\partial_0\beta+\{\beta,{\alpha}\},$$
where the Poisson brackets are defined with respect to the measure $w$
\eqn\pois{\{A,B\}= {\epsilon^{rs}\over w(\sigma)}
\partial_rA(\sigma)\partial_sB(\sigma).}

In order to follow the decoupling limit more carefully we would like
to re-introduce $P_-$ and $M_p$ into the Lagrangian. This is
determined by dimensional analysis and longitudinal boost
invariance. At this point we will also switch to an open world volume of 
the form (a 2-disk of area 1)$\times$ time 
(we also set the measure $w$ to 1). The light cone quantization
for the open membrane is discussed at greater length in \opnlc. The
action we obtain is:
\eqn\dimlag{{\cal L}_0=
{P_-\over 2}(D_0X)^2-{M_p^6\over 4P_-}(\{X^a,X^b\})^2}

Finally we would like to insert the topological term that comes from
the coupling of the boundary to the 3-form field strength. Although
it is a boundary term it is more convenient to write it as an total
derivative integrated over the bulk of the field
\eqn\toptrm{{\cal L}_1=\{X^a,X^b\}H_{+ab}}
This is of course a special case of the topological term \brga\brgb\
\eqn\brgtop{\int_{\partial M^3} d\tau d\sigma H_{\mu\nu\rho}
X^\mu{\dot X}^\nu{X'}^\rho} after using the lightcone gauge condition \lccb. 

One should, however, be very careful how one uses this action. The reasons 
are familiar:
\item{1.} Strictly speaking, this action corresponds to the first 
quantized theory (like the cylinder in string theory). The problem is that, 
as mentioned before, the theory is strongly interacting and hence 
single and multi membrane states mix strongly.
\item{2.} There are still the usual IR problems associated with membranes 
( in the form 
of thin long low-energy spikes extending far from the membrane).
\item{3.} As a 2+1 field theory, it is not clear how to make sense of it.

These problems are the same as for the closed membrane, and we expect
that a matrix regularization, and a proper interpretation of it, will
solve them in the same way that the BFSS matrix model ``solves'' the
problems of the membrane theory of \clmem. However, even from the
perspective of the BFSS model, the action for the membrane is useful
for some questions. For example, constructing membranes in Matrix
theory essentially generates this Lagrangian from the matrix partons
\base. Hence we can also hope that some qualitative understanding of
the dynamics of membranes can be achieved from the Lagrangian
\dimlag+\toptrm. 

We will therefore analyze this Lagrangian semi-classically to obtain
some qualitative understanding of its dynamics. 
A static extremum of this action satisfies
\eqn\eqmblk{ \{X^a,\{X^a,X^b\}\}=0 }
with the boundary condition
\eqn\bndr{ -{M_p^6\over P_-}\{X^a,X^b\}\partial_{\vert\vert}X_b + 
2H_+^{ab}\partial_{\vert\vert}X_b=0.}
The boundary condition is derived by
requiring that no energy leaks from the membrane's boundary. The rate
of energy loss is proportional to $D_0 X^a$ times the LHS of equation
\bndr. Requiring that it is zero for all values of $D_0 X$ then gives 
\bndr. Note that the derivatives are parallel to the boundary rather 
than transverse to it. This is so because of the unusual form of the 
gradient energy.

We would like to simplify this set of equations, but in a way that
still captures the interesting physics. The most relevant degrees of
freedom for an extended object with a dipole charge is of course, its
area and orientation, and we will focus on these degrees of
freedom. We will therefore take the membrane to be planar and of a fixed
shape. More precisely we will assume that the 4 coordinates transverse
to the light cone are of the form
\eqn\crdfrm{X^i(\tau,\sigma_1,\sigma_2)=\Sigma_{j=1,2}X^i_j(\tau)\sigma^j.}
This is in the spirit of \lennydip\ where the degrees of freedom of
the dipole in non-commutative geometry are encoded in its length and
orientation.

This ansatz automatically satisfies the bulk equation of motion (we
will return to the boundary conditions shortly). The issue now is that
the minimum of the potential is degenerate. This manifold of
degenerate vacua is what interests us the most. In a situation like
this the wave function of the ground state of the system is
approximately a uniform wavefunction on the manifold of degenerate
vacua\foot{Because we are in finite volume, the system is quantum
mechanics rather then a quantum field theory with different
superselection sectors.}. In our case the remarkable thing is that
this submanifold will be closely related to a certain submanifold of
the ADHM space, and indeed the ground state of quantum mechanics on
the ADHM manifold will be concentrated around the latter.

For now let us proceed to study the manifold of degenerate vacua.  It
is easy to see why there are degenerate minima. Completing to a square
the potential is
\eqn\potsqr{{M_p^6\over 4P_-}
\Sigma_{a,b}\bigl( X^a_1X^b_2-X^a_2X^b_1-{2P_-\over M_p^6}H_+^{ab}\bigr)^2}
For the potential to have a single non-degenerate minimum, then all
 the terms vanish independently. It is easy to see that this is
 impossible to achieve.  For example if we pick (without loss of
 generality) $H_{+12}=H_{+34}\not=0$ then to satisfy \bndr\ we need to
 set, for example,
$$\partial_{\sigma_1}X^1,\partial_{\sigma_2}X^2\not=0 
					\rightarrow \{X^1,X^2\}\not=0$$
$$\partial_{\sigma_1}X^3,\partial_{\sigma_2}X^4\not=0
					\rightarrow \{X^3,X^4\}\not=0$$
but then we run into the problem that 
$$\{X^1,X^4\},\{X^2,X^3\}\not=0$$ as well.

Characterizing the manifold of degenerate vacua is not difficult. 
Switching to an
$SU(2)_L\times SU(2)_R\sim SO(4)_{transverse}$ notation $X^{ai}\rightarrow
X^{\alpha{\dot\alpha}i},\ i=1,2$ and regarding the pair of $X^i$ as $2\times
2$ matrices (in the $\alpha\dot\alpha$ indices) the potential is
\eqn\redpot{
{\vert X^1X^{2\dagger}-X^2X^{1\dagger}-H_+ \sigma^2\vert} ^2+ {\vert
X^{1\dagger}X^2-X^{2\dagger}X^1 \vert}^2.} (the $a,b$ indices of $H_+$
are encoded in the matrix $\sigma^2$). It is now easy to minimize the
potential. Using an $SU(2)_R$ (the one broken by $H_{+ab}$) we can
minimize the first term by bringing the matrix
$X^1X^{2\dagger}-X^2X^{1\dagger}$ to be parallel to $H_{+ab} =
H_+ \sigma^2$. $SU(2)_L$ then spans the degenerate vacua. Up to
irrelevant constants the vacuum manifold is therefore given by
\eqn\sol{ \{X^a,X^b\}_{self-dual}\propto H_{+ab}} 
$$ (\{X^a,X^b\}_{anti-self-dual})^2 \propto \vert H_+ \vert^2$$
Finally, in regards to the boundary conditions, one can show that
every configuration along the minima satisfies \bndr, in fact the
equations for the minima (in terms of $X^{1,2}$) are precisely
\bndr. We will verify the predictions \sol\ in the discrete light cone
model in sections 3 and 4.

Another way of understanding this degeneration is the following. In
the case of NCYM the toplogical Lagrangian is ${\dot
X}^iB_{ij}\delta^j$ where $\delta^j$ is the size of the dipole in the
j direction \lennydip\sw\zy. This gives the relation
$P_i=B_{ij}\delta^j$, which can be inverted (if $B$ is degenerate we
can not restrict ourselves to the topological term) to yield a
specific dipole direction as a function of the momentum. In the case
of a 3 form the topological Lagrangian is ${\dot X}^i H_{ijk} V^{jk}$
where $V^{jk}$ is the volume in the j-k plane. This can not be
inverted to yield a specific membrane orientation as a function of the
momentum, i.e., there is a degeneracy.

Two comments are in order. The first is that we have dropped a
significant amount of information in this approximation. It should
not, however, be difficult to reinstate it. The 2nd is another
precursor to what is to come. The readers familiar with the D0-D4
system will realize that the self-dual part of the potential \potsqr\
closely resembles the structure of the F/D-term constraints of the
D0-D4 system to which we turn in section 3.

\subsec{Another Approach to Open Membranes}

We have discussed so far an approach to open membranes in which one
takes the world volume to be a disk. This disk replaces the sphere in
the case of the closed membrane. One then needs to consider
reparametrizations which keep the boundary of the disk fixed, i.e,
only a subgroup remains. This approach is discussed in \opnlc\havgrp.
However, in the DLCQ description that we will use \abs\ (which we will
review in the next section) the entire $U(N)$ gauge symmetry is kept,
i.e., the reparametrization group of the sphere is
retained\foot{Although for every finite N we see only a subgroup of
it.}. We would therefore like to keep the topology of a closed sphere
even when discussing the world sheet of open membranes.

This may sound impossible but is actually very familiar - a similar
thing happens in string theory. There the insertion of any vertex
operators on the world sheet is equivalent to adding a boundary to the
world sheet. By conformal invariance such an insertion is equivalent
to an infinite tube, and the particle content of the vertex operator
is described by boundary conditions at the end of that tube. But as is
clear, even though the vertex operator is an implicit boundary, the
correct symmetry group is the conformal symmetries of the sphere,
except that the vertex operator transforms under these symmetries.

A very similar thing will happen here. One can make the closed sphere
into an open sphere by inserting an ``impurity'' at some point along
the sphere. This impurity will allow the fields to be non-smooth
around this point, which will in effect make it into an open disk. As
in the case of string theory, these impurities will transform under
the reparametrizations of the sphere, i.e, under the $U(N)$ gauge
symmetry. The readers already familiar with the model in \abs, will
guess correctly that these are fundamental hypermultiplets.

In the remainder of this subsection we will try and quantify how many
``impurity'' degrees of freedom are required in order to make a closed
surface into an open one. This is done in preparation for the next
section where we will present the discrete light cone description of
this model, and where it will become clear what these impurities are.

Instead of 4 real coordinates, we use 2 complex coordinates
$X,{\tilde X}^*$, which are $SU(2)_R$ doublet ($SU(2)_L$ mixes $X$ and
$\tilde X$). The potential is
\eqn\potmat{\int d^2\sigma Tr \Biggl(
\matrix{\{X,X^*\}+\{{\tilde X},{\tilde X}^*\} & -2\{X,{\tilde X}\}\cr
	2\{X^*,{\tilde X}^*\}	& \{X,X^*\}+\{{\tilde X},{\tilde X}^*\} }
\Biggr)^2}
Suppose we allow the fields to be discontinuous at one point, i.e,
as one approaches the point from different directions, the limiting
value of the fields may be different. One typically does not allow
this because of energetic reasons. Suppose we cut off a small circle
of radius $\epsilon$ around the point, and we allow the fields to be
arbitrary around the hole. In this case, the generic behavior of
$X,{\tilde X}$ around a singularity of the type which interests us is
$$\matrix{ X&=f_0(\theta)+f_1(\theta)r+f_2(\theta)r^2+...\cr {\tilde
X} &=g_0(\theta)+g_1(\theta)r+g_2(\theta)r^2+.  }$$ which gives a
Poisson bracket which diverges as $1\over r$. Hence the divergence
behaves like\eqn\divlvl{\int_\epsilon rdr {1\over r^2} d\theta
f(\theta),} where $f$ is some functional of $f_0,g_0,f_1,g_1$. In
order to allow for the fields to fluctuate in a generic fashion we
would like to eliminate this divergence. This can easily be achieved
by inserting some dynamical degrees of freedom at the singularity,
such that these exactly cancel the singular part of the Poisson
bracket. The action will therefore now be
\eqn\modpot{\int d^2\sigma Tr\Biggl(
\matrix{\{X,X^*\}+\{{\tilde X},{\tilde X}^*\} & -2\{X,{\tilde X}\}\cr
	2\{X^*,{\tilde X}^*\}	& \{X,X^*\}+\{{\tilde X},{\tilde X}^*\}} 
+\Delta(\sigma) {1\over \epsilon} L_{2\times 2} 
\Biggr)^2,}
 where $\Delta$ is some distribution localized around the specific
 point which we excise from the worldvolume.

Our task is to quantify what is the minimal set of degrees of freedom
in $L$ such that we will remove the $log(\epsilon)$ singularity in the
energy. The matrix of Poisson brackets is clearly 2 by 2
anti-hermitian matrix in the adjoint of $SU(2)_R$, hence we will
require the same from $L$. Such matrices can be parameterized by a
single vector which is a doublet of $SU(2)_R$, where $L$ is their
bi-linear. Hence we see that in order to open a hole in the world
sheet we need a single doublet of $SU(2)_R$. If we denote this vector as 
$(Q_1,Q_2)$ then 
\eqn\frml{L=\biggl(\matrix {Q_1^*\cr Q_2^*}\biggr)(Q_1,Q_2)
	   -\biggl(\matrix {Q_2\cr -Q_1}\biggr) (Q_2^*,-Q_1^*)} The
values that the $Q$ impurity fields will take in order to obtain a
finite energy are given by
\eqn\imptcnstrnt{lim_{\sigma\rightarrow\ excised\ point}
\biggl( \matrix{ \{X,X^*\} - \{{\tilde X},{\tilde X}^*\} + 
				Q_1Q_1^* -Q_2Q_2^* \cr -2\{X,{\tilde
X}\} +2Q_1^*Q_2}\biggr)=0} Note again these these look precisely like
F/D-terms for a ${\cal N}=8$ system. After discretizing by matrices,
these will in fact become exactly that.  The fact that these are point
like impurities leads to these becoming vectors of $U(N)$ after
discretization. We will encounter precisely such fields in the
discrete light cone quantization, to which we move next.

\newsec{The Resolved (2,0) Matrix Model}

\subsec{Review of the Model}

The discrete light cone quantization of the (2,0) field theory with a
large 3-form turned on is given in \abs, where it appeared as a
natural generalization of the DLCQ of the (2,0) field theory \abkss.
The model for the latter is quantum mechanics on the moduli space of
Instantons, and turning on a 3-form field (in the decoupling scaling
described above) corresponds to resolving this manifold. The model
when the field strength is turned on is therefore actually under
better control than that of the (2,0) SCFT.

For $k$ M5-branes and momentum $p_-=N/R$, the moduli space of
Instantons in question, $\mnk$, is the moduli space of $N$ instantons
in a $U(k)$ group. The deformation which takes us from the (2,0) CFT
to the light-like ``non-commuative'' theory corresponds to turning on
a non-zero FI term, which makes the space smooth. We will begin by
discussing the unresolved model (of the (2,0) CFT) and then discuss
the resolution.

\medskip
\noindent{{\it 3.1.1} The ``commutative'' model} 
\medskip

A simple concrete description of our theory is as the Higgs branch of
a $U(N)$ gauge theory with 8 supercharges, an adjoint hypermultiplet
(consisting of 2 complex adjoint field $X,{\tilde X}$), and $k$
fundamental hypermultiplets (consisting of $Q_i, {\tilde Q}^i,\
i=1..k$ in the fundamental (antifundamental) of $U(N)$. We will also
denote these occasionally as $Q^\alpha_i,\ i=1,..k,\ \alpha=1,2$).  The
Higgs branch is parameterized by the values of these fields, subject to
the vanishing of the F/D-terms:
\eqn\Dconstraints{[X,X^\dagger] + [\tX,\tX^\dagger] + Q_i Q_i^\dagger
- ({\tilde Q}^i)^\dagger ({\tilde Q}^i) = 0}
and
\eqn\Fconstraints{[X,\tX] + Q_i {\tilde Q}^i = 0,}
modulo $U(N)$ gauge transformation. The total (real) dimension of
this space is $4Nk$, and it is a hyperK\"ahler manifold. The space is
also equipped with a natural hyperK\"ahler metric which is the
restriction of the flat metric on the linear space to $\mnk$.

The global symmetries of the theory are $SU(2)_R \times SU(2)_L \times
Spin(5) \times U(k)$. The first two factors form an $SO(4)$ which
correspond to the rotation of the 4 direction transverse to the
lightcone coordinates (inside the 5-brane). Spin(5) is an R-symmetry
of the (2,0) CFT (which can be understood geometrically as rotations
transverse to the 5-brane). The last factor, which is a global flavor
symmetry in the quantum mechanics, is believed to have a 6D dimensional 
interpretation as a global remnant of
the gauge symmetry of the (2,0) CFT, after some gauge fixing when
going to the light cone. The supercharges in the quantum mechanics are
in the $\bf(2,1,4,1)$ representation of this group,
and the fields described above transform as follows:
\eqn\representations{\matrix{
&U(N)&SU(2)_R&SU(2)_L&Spin(5)&U(k)\cr
X_H&\bf{N^2}&\bf{2}&\bf{2}&\bf{1}&\bf{1}\cr
\Theta_X&\bf{N^2}&\bf{1}&\bf{2}&\bf{4}&\bf{1}\cr
Q_H&\bf{N}&\bf{2}&\bf{1}&\bf{1}&\bf{k}\cr
\psi_Q&\bf{N}&\bf{1}&\bf{1}&\bf{4}&\bf{k},\cr}}
where $X_H$ denotes the scalars $X$ and $\tX$, and $Q_H$ denote the
scalars in the fundamental $Q$ and ${\tilde Q}^*$ (and $\Theta_X$ and
$\psi_Q$ denote their superpartners).  

\medskip
\noindent{{\it 3.1.2} The ``Non-commutative'' case}
\medskip

We pass from the ordinary (2,0) CFT to the theory with light like
(2,0) ``non-commutativity'' by turning on a field strength $H_{+ab}$
where $a,b$ are indices in the 4 coordinates transverse to the light
cone coordinates. The self-duality relation then amounts to requiring
that $H$ is self-dual in the indices $a,b$. Its quantum numbers are
therefore such that it transforms under $SU(2)_R$ and is invariant
under all the other symmetries of the model. 

Since this model has 8 supercharges, corrections to
it are very restricted. Fortunately, there is a simple deformation
with precisely the right quantum numbers, which is a deformation of
the model by turning on a FI term in the $U(1)$ part of $U(N)$. This
changes the F/D-term constraints (and breaks the $SU(2)_R$) but is
otherwise a fairly mild deformation of the model above. By an
$SU(2)_R$ rotation we can bring the F/D-terms to the form
\eqn\modcon{[X,X^\dagger] + [{\tilde X},{\tilde X}^\dagger] + Q_i Q_i^\dagger
- ({\tilde Q}^i)^\dagger ({\tilde Q}^i) = \zeta}
$${[X,{\tilde X}] + Q_i {\tilde Q}^i = 0}.$$

It is also easy to figure out the precise relation between $\zeta$ and
$H$. This relation is uniquely determined by dimensional analysis and
longitudinal boost invariance to be (up to a constant which will not
be important to us)
\eqn\rel{\zeta\propto{{H\over {RM_p^6}}}}
For a more detailed discussion the reader is referred to \abs\agm.

\subsec{The Open Membrane Interpretation}

We have discussed before how to generate open membranes from closed
ones by the addition of impurity degrees of freedom on the
worldvolume. In this section we will relate this discussion to the
discrete light cone model above. Roughly, keeping the topology of a
closed sphere corresponds to keeping $U(N)$ as a gauge symmetry and
keeping the adjoint matter fields. The addition of impurity degrees of
freedom corresponds to the addition of the hypermultiplets $Q$. For
the purposes of this section we will work prior to decoupling, i.e.,
we will not impose that the F/D-terms are precisely zero, but rather
one pays a finite amount energy for their violation. This will have
the effect of allowing the membranes to fluctuate more freely and it
will be simpler to analyze their structure. We will also not be
turning on an $H$ field.

It is known that the procedure \clmem\ for discretizing closed membranes
actually results in the BFSS matrix model.
\eqn\matrep{ X(\tau,\sigma)\rightarrow X(\tau)_{N\times N}}
 $$\{A,B\}\rightarrow {1\over N}[A,B]$$
 $$\int d^2\sigma A \rightarrow {1\over N}Tr(A)$$
 $$reparam.\ \rightarrow U(N)\ gauge\ trans.$$
These relations continue to hold pretty much as they are in our case.
  
Next we would like to substantiate the role of the hypermultiplets $Q$
as the impurities which open the surface:
\item{1.} We examined before what quantum numbers these impurities are 
required to have, and it turned out that a doublet of $SU(2)_R$ did
the job. These are precisely the quantum numbers of the $Q$'s.
\item{2.} We can examine related occurrences of the hypermultiplets $Q$ in 
matrix models, and see if they correspond to making closed world
volumes into open ones. This is indeed the case. Such impurities are
responsible, for example, for making closed strings into open strings.
For our purposes any Matrix model of a background with a D-brane would
do, and we would focus on the case of the D4-brane, which is discussed
in \mdfour. This case is of direct relevance to the M5-brane case
since it is its double dimensional reduction, and turning closed
strings into open ones is precisely turning closed membranes into open
ones.

\item{} The Matrix model for IIA in the presence of a D4-brane is readily
derived from that of the Matrix model of M-theory in the presence of a
5-brane. One goes to the IIA limit by compactifying on a circle, and
the model is now the 1+1 ${\cal N}=16$ SYM on a circle with the
addition of ${\cal N}=8$ hypermultiplets localized at discrete points
along the circle, $\sigma_i$. The theory away from the impurities
has 8 scalar fields out of which 3 parameterize fluctuations parallel
to the D4-brane, and the $SU(2)\sim SO(3)$ symmetry that rotates them
is nothing but the R-symmetry of the model. The F and D-term
constraints of this system are therefore:
\eqn\impu{D_\sigma X^a-\delta (\sigma-\sigma_i) 
Q^\alpha\sigma^{a\beta}_{\alpha}Q^*_\beta=0} Open strings are now
generated when the hypermultiplets $Q$ obtain a VEV. In this case the
$X$ variables can jump between the two sides of the impurity. In fact,
if we solve for $Q$'s in terms of the $X$'s then any jumps are
allowed. The value of the coordinate $X$ as one approaches the
impurities from above or below corresponds to the position of the end
of the string, and the process by which the value of $Q$ changes from
being zero to being non-zero correspond to a closed string splitting
into open strings. We are less interested here in the details of this
process but the main lesson that we would like to draw is that the
generation of holes in the worldsheet is a dynamical process written
in terms of world sheet variables, i.e., the hypermultiplets get a VEV
and ``cut the world sheet''.  Since we have similar hypermultiplets in
the ADHM sigma model, the interpretation is clearly similar. 
\item{3.} Finally, one can examine the form of the $Q$'s within the context 
of fuzzy spheres. Let us focus on one of the
elements in the $SU(2)_R$ doublet, say $Q^1$, which we will denote by
$Q$ and take to be a column vector of $SU(N)$. The schematic form of
the D-term constraints is
\eqn\schmd{[X_i,X_j]+QQ^\dagger + .... =0}
We are accustomed to thinking about the commutator $[X,X]$ as
measuring the volume of the membrane, and this suggests $QQ^\dagger$
is also a volume. More precisely, we would like to identify it as a
function on the sphere which has a support of volume $1/N$, i.e, the
smallest allowed volume (in this sense it should be considered as a
``point'' on the fuzzy sphere). Furthermore, since we are not allowed
to have any more detailed information on such a small volume, it
should be interpreted as a fixed function on such a volume.

\item{}As supporting evidence for this interpretation, one can consider the
following simple example. Suppose we are interested in a function that
is 1 on a strip of width $1/\sqrt{N}$ around the equator. 
When integrating such a function with any other function which
vanishes close to the equator, then the result is zero. Embedding the fuzzy sphere in $R^3$ with coordinates $z_{1,2,3}$, then any function that
vahishes near the equator is of the form $z_3\times P(z_1,z_2,z_3)$
where $P$ is an arbitrary function of the $z$ variables. Hence we
require that the matrix corresponding to the equator indicator
function M will satisfy
\eqn\trreq{Tr(z_3P(z)M)=0 \ for\ all\ P\rightarrow Z_3M=0}
Taking $Z_3$ to be the diagonal $(\sqrt{N},....-\sqrt{N})$ we see that
$M$ is of the form $QQ^\dagger$ with
\eqn\qval{Q^\dagger=(0,0,... 1_{N\over2},0,... 0)}

\item{}Hence we see that the $Q$ is a degree of freedom associated with a
minimal area insertion on the sphere - a slightly thickened point or a
line. We have chosen here a matrix M which corresponded to the equator
but it is not difficult to see that there are choices which correspond
to, for example, the poles of the sphere\foot{The choices of different
geometries for the minimal area patches probably corresponds to
different orbits of the $SU(2)$ isometry of the sphere in the $N$'th
dimensional representation.}.  More precisely, since the smallest
volume on the fuzzy sphere is $1/N$ of the volume of the sphere, and
we have at most $k$ $Q$ vectors, the maximal excised volume if
$k/N$ times the volume of the sphere, which is negligible in the
$k<<N$ matrix theory limit.  We should think about it as an
``anti-surface''. By the insertion of such an impurity we allow the
surface to have a boundary, which is no other then the boundary
between the minimal impurity surface and the rest of the sphere. Hence
we are able to generate open membranes from closed ones.

\subsec{Measuring Open Membrane}

This also leads us to the more precise way of measuring the surface of
the membrane. We have use the fact before that $[X^i,X^j]$ provides us
with information about the volume in the $i-j$ plane. The problem is,
of course, that if we want to get a number for the volume the we face
the obvious problem that $Tr([X^i,X^j])=0$. In the usual applications
of Matrix theory this does not bother us, since the volume comes with
a sign (i.e, its the source for a charge) and the two orientations of
the closed compact membrane cancel each other, yielding the correct
result that the signed volume is zero. However, for the case of the
open membrane it is a problem since an open stretched compact membrane
can have a net signed volume.

One way out is, as for closed membranes, to work with dipoles and
higher moments. This, however, is unsatisfactory since the volume is a
well defined concept and we would like to be able to measure it. It
also turns out that in specific cases (discussed later) naively
computing the moments gives results which are difficult to
interpret. There is fortunately a simple resolution, along the lines
of the discussion above, which matches our expectations (in the large
$N$ limit).

Recall that we have divided our sphere into two parts where the
smaller part is actually excised, and is indicated by where the
$Q,{\tilde Q}$ vectors are supported. By the volume of the membrane in
spacetime we actually mean the volume that is covered by the rest of
the sphere. Therefore if we want to measure the volume we actually
need to integrate over the sphere except the excised patch. Denoting
by ${\cal A}$ the excised patch on the sphere, and by $ind({\cal A})$
a function that is 1 on that patch and 0 elsewhere, then the
integration we are interested in is actually
\eqn\volintg{\int d^2\sigma \{X^1,X^2\}(1-ind({\cal A})).}
which has the immediate generalization to the non-commutative case
\eqn\volncg{Tr\bigl([X^1,X^2](1-P)\bigr)}
where $P$ is the matrix that corresponds to the indicator function,
i.e., it is a projection operator which projects out the subspace
spanned by $Q,\tilde Q$. Note that typically this is a two dimensional
space, but it may degenerate to a one dimensional space in special
points on the moduli space. In the non-degenerate case the matrix is given by 
\eqn\prj{P={QQ^\dagger \over \vert Q\vert^2} + 
	   { {\tilde Q}^\dagger{\tilde Q}\over \vert {\tilde Q}\vert
^2}.}  This is the first step in the definition but it is not complete
because we still need to specify how to compute
\eqn\morcmp{ \int d^2\sigma \{X^1,X^2\} (1-ind({\cal A})) f(\sigma_1,\sigma_2)}
for an arbitrary $f$. The issue is to specify the insertions of $1-P$
for the general case. To solve this we would like to find a matrix
form for $\{X^1,X^2\}$, which already includes the $1-P$ projection,
and would be suitable for integration at with any function. At this
point we will motivate such a guess, and will check it in a later
section\foot{More tests are clearly needed since we will check only a
few very special cases.}. There it will be clear that in order to
obtain sensible results it is necessary to insert the projection
operator $1-P$.

Recall that in \base\ for the case of a closed membrane, the charge
density $[X^1,X^2]$ was obtained by a computation analogous to a
computation of the central charge density in BPS formulas via commutators
of charges and currents. In order for the central charges to be real,
then $[X^1,X^2]$ has to be anti-hermitian (the analogue of imaginary
for matrices). The charge density that we will define will still have
this property. Hence we would like to modify the expression
$[X^1,X^2]$ by $1-P$ in a way which preserved anti-hermiticity. The
natural guess is
\eqn\crgdns{charge\ density\ = (1-P)[X^1,X^2](1-P).}
We will shortly provide some checks on this expression for specific
discretized open membrane configurations.

Finally let us examine the volume given by this prescription and
compare them to the volumes we estimated before.
\eqn\vla{Tr\bigl([X,{\tilde X}](1-P)\bigr)=0}
\eqn\vlb{Tr\bigl(([X,X^\dagger]+[{\tilde X},{\tilde X}^\dagger])(1-P)\bigr)
\propto
(N-2)\zeta} i.e, these are the same volumes as we suggested before in
\sol, in the large N limit.

\subsec{An Example: Equations of motion}

We have seen before that the finite energy configuration space for the
membrane is 
\eqn\blkeq{\{X^a,\{X^a,X^b\}\}=0.}
For all other configurations one has to pay an energy of
proportional to $M_p$, i.e., they are not allowed. We would like to
verify this relation in the Matrix model.

This will also serve to check the prescription \crgdns, that the
membrane is actually described by the $X$ matrices projected by
$1-P$. We are interested in the product of $\{X^a,\{X^a,X^b\}\}$
restricted to the non-excised volume of the sphere. Hence a guess for
this product would be the symmetric product
\eqn\plntry{ (1-P)[X^a,[X^a,X^b]](1-P)}
on the ADHM moduli space. We will evaluate
\eqn\spcpln{ [X^a,[X^a,X^1]]=}
$$=-{1\over 4}[X-X^\dagger,-QQ^\dagger+{\tilde Q}^\dagger{\tilde Q} ]
+{1\over 2}[{\tilde X}^\dagger,Q{\tilde Q}]
-{1\over 2}[{\tilde X},{\tilde Q}^\dagger {\tilde Q}]$$
where we have used the F/D-term.
This clearly satisfies
 \eqn\plnfin{(1-P)[X^a,[X^a,X^1]](1-P)=0}
which is the expected result. 

\newsec{Some examples}

We would now like to discuss the open membrane interpretation of some
points in $\mnk$, as a check on the formulas we derived above. For
most of the $\mnk$ explicit solution for the ADHM constraints are not
known, but some special cases are known.  The case that we will focus
on is the one that corresponds to a single 5-branes, i.e.,
instantons in $U(1)$. The manifold of two such instantons is very
familiar and we will discuss it at length as our basic check. Then we
check the validity of the formulation for an arbitrary number of
instantons on a configuration of a large round open disk, and finally
we will briefly discuss well separated multi-membrane configurations
as a check that the model factorizes properly.

\subsec{The $P_-=2/R$}

We begin by discussing the case of two instantons in a $U(1)$ gauge 
field. Before turning on the blow-up
parameter the moduli space is
\eqn\mdlzro{{\cal M}_{2,1,\zeta=0}=R^4\times (R^4/Z_2)}
The FI term acts as a blow-up of the $Z_2$ singularity and the 2nd
component is replaced by a smooth ALE space.

The $R^4$ part has the obvious interpretation of the center of mass
coordinate for the entire system. The $R^4/Z_2$ part, before blowing
up, is also easy to interpret - it describes the relative position of
two identical particles (each carrying one unit of momentum). The
metric on the relative position component is flat at any non-zero
separation of the particle, which corresponds to the fact that the
single (2,0) multiplet theory, without an $H$ field, is a free theory.

The interpretation as two particles is not the entire story.  Even
before blowing up one can ask where is the single particle state with
two units of momenta. There is a natural candidate for it, which is
the harmonic wave function dual to the shrunken cycle at the origin of
(defined by a limiting process as we shrink the cycle, and defines a
singular form at the end). But because the theory is free, it is up to
us whether or not we include such a state. Since the two $p_-=1/R$
particles can not dynamically combine to make the $p_-=2/R$ state, there
is no dynamical necessity for including it at this point.

By turning on a non-commutativity parameter we introduce interactions
into the theory. It is now necessary that the Hilbert space includes
the single particle $p_-=2/R$ state. Indeed, ${\cal M}_{2,1}$ has now
been smoothed out, and the aforementioned harmonic form is a smooth
well behaved form on this space. Hence it is automatically included in
the Hilbert space of the theory (which the Hilbert space of $L^2$
harmonic forms on the space) and there is in fact no natural way of
separating it out.  The state which interests us the most is precisely
this single particle $p_-=2/R$ state. The reason is that it is this
type of states which are relevant in the limit $N\rightarrow\infty,\
E_{DLCQ}\propto 1/N$ (which is necessary in order to obtain
non-compact space-time results from a discrete light cone quantization
procedure). This is similar to the statement that in the BFSS model,
the relevant excitations are the bound states of D0-branes with a
finite fraction of the null momentum.

After blow-up this form is perfectly regular and manageable. Since it
corresponds to a single particle state it is actually as close as we
can get to the state of a single membrane (up to the caveats mentioned
before). Because we have put all the momentum in a single membrane, it
is also the largest membrane in the $p_-=2/R$ sector, which simplifies
the analysis.

Fortunately the Harmonic form corresponding to this states is very 
familiar. In the resolved space we have blown-up the
singular point into a sphere of some finite radius set by $\zeta$. The
harmonic form is localized near this two sphere and decays (like a
power law) as we move away from the sphere (and from what used to be
the origin of the space before blow-up). To have an interpretation
of a single open membrane we clearly need to stay close to this
sphere because as we go away the interpretation as two separate
$p_-=1/R$ particles becomes better. The fact that there is a smooth
transition from one interpretation to another, and that the wave function
is, strictly speaking, supported at both is nothing but a manifestation 
of the single-membrane/multi-membrane mixing. 
 
Therefore, we would like to study configurations along this 
minimal sphere and see whether they agree with out
expectations from the previous sections. 
The form of the ADHM matrices on this cycle are given in \nikita\newk, 
and are
\eqn\sprmat{     X=\biggl(\matrix{ 0 & p_1 \cr 0 & 0}\biggr),\ \ 
	{\tilde X}=\biggl(\matrix{ 0 & p_2 \cr 0 & 0}\biggr),\ \
		 Q=\sqrt{2\zeta}\biggl( \matrix{ 0 \cr 1}\biggr),\ \
		{\tilde Q}=0}
where $p_1$ and $p_2$ are complex numbers satisfying the constraint
$${\vert p_1 \vert}^2+{\vert p_2 \vert}^2=\zeta.$$ 
The 4 real coordinate matrices are then easy to derive ($X=X_1+iX_2,\ {\tilde
X}=X_3+iX_4$) and, using formula \volncg, give us the following volumes:

The self-dual volume:
\eqn\sldvl{\matrix{ 	V_{12}+V_{34}=&{1\over2}\zeta   ,\cr
			V_{13}-V_{24}=&0   ,\cr 
			V_{14}+V_{23}=&0			}}
and the anti-self-dual volume:
\eqn\ansldl{\matrix{	V_{12}-V_{34}=&{1\over2}
					(\vert p_1\vert ^2 - \vert p_2
			\vert^2),\cr V_{13}+V_{24}=& {1\over 2i}
			(p_1p_2^*-p_1^*p_2) ,\cr V_{14}-V_{23}=&
			-{1\over 2} (p_1p_2^*+p_2p_1^*) }} 
The right hand quantities can be written as
$$\propto (p_1,p_2) \sigma^i (p_1,p_2)^\dagger$$
Both of these match precisely our expectations from section 2.

Next we would like to check the situation of a pair of well
separated particles, each with momentum $p_-=1/R$. As we explained 
above, this is the useful approximate description far away from the
origin of the $R^4/Z_2$. We will see that in this regime the 
insertion of the projection operator $1-P$ (as in 
\volncg) is necessary in order to compute some quantities,
(although it does not solve all the puzzles in other quantities).

The general solution for 2 instantons in $U(1)$ was given in
\newk. The solution is (neglecting the center of mass)
\eqn\sepinst{
X={z_0\over2} \bigl(\matrix{ 1&\sqrt{{2b\over a}}\cr
			  	   0&-1 }\bigr),\ 
{\tilde X}={z_1\over2}\bigl(\matrix{ 1&\sqrt{{2b\over a}}\cr
			  	   0&-1 }\bigr)}
$$Q=\sqrt{\zeta}\bigl(\matrix{\sqrt{1-b}\cr \sqrt{1+b}}\bigr),\ \ 
{\tilde Q}=0$$
where
\eqn\mrinf{
a={ {\vert z_0\vert}^2+{\vert z_1 \vert}^2\over 2\zeta},\ \ b={1\over
a+\sqrt{1+a^2}}} When $z_0$ and $z_1$ are taken to infinity this
corresponds to two well separated particles (each blown up into a
small membrane).

We would like to check how our procedure fares in computing the
position of the particles. We will compare the results with and without 
the insertion of the projection operator, i.e, for example,
\eqn\exmn{Tr \bigl(([X,X^\dagger]+[{\tilde X},{\tilde X}^\dagger]) X^{2l} 
\bigr)\ \ 
\ vs.}
$$Tr \bigl((1-P) ([X,X^\dagger]+[{\tilde X},{\tilde X}^\dagger])(1-P) X^{2l} 
\bigr)$$
where we have used the volume elements (the commutator) parallel to $H_+$ to 
indicate where the particles are (the commutator is non-zero there, 
indicating that there is a small piece of open membrane there). 
Computing the trace without the projection operator might
also appear to be a natural extension
of the computation for the closed membrane.

Evaluating these expression to the leading order 
$1/z^2=1/(\vert z_0 \vert^2+ \vert z_1 \vert ^2)$ we obtain that 
the first expression is
$$Tr \biggl( \bigl(\matrix{ 
    4\zeta^2/\vert z\vert ^2 & -4\zeta \cr
    -4\zeta & -4\zeta^2/\vert z\vert^2 } \bigr) 
   { \bigl( {z_0\over 2}\bigr)}^{2l} \bigl( \matrix{ 1&0\cr 0&1} \bigr)
\biggr)\sim {\vert \zeta\vert^2 \over \vert z_0 \vert ^2 + 
					\vert z_1 \vert ^2} 
{\bigl( {z_0\over 2} \bigr)}^{2l}$$
and the 2nd is 
$$Tr \biggl( \bigl(\matrix{1/2& -1/2\cr -1/2&1/2}\bigr)
\bigl(\matrix{ 
    4\zeta^2/\vert z\vert ^2 & -4\zeta \cr
    -4\zeta & -4\zeta^2/\vert z\vert^2 } \bigr) 
    {\bigl( {z_0\over 2}\bigr)}^{2l} \bigl( \matrix{ 1&0\cr 0&1} \bigr)
\biggr)\sim {\bigl({z_0\over2}\bigr)}^{2l} $$
where the 1st matrix in 2nd expression is the projection operator
$$1-P \rightarrow 1- {1\over 2}\bigr(\matrix{1&1\cr
1&1}\bigr),$$ using \sepinst. It is clear that without inserting the projection
operators we obtain results which are difficult to interpret, whereas
with the insertion of the projection operators we precisely get the
expected results.

On the down side, even with the projection operator not all expression
are easy to interpret. For example one can evaluate the volumes in the
different directions.  Of the self-dual volumes, only
$Tr\bigl((1-P)([X_1,X_2]+[X_3,X_4])\bigr)$ is non-zero, as it should
be. However, the anti-self-dual volumes are also not zero. Unlike when
discussing a single membrane, this result is not what one expects,
although only applying some indirect arguments. When the two membranes
are far apart their anti-self-dual volumes are uncorrelated (the
self-dual-volume are correlated with the external field, and therefore
parallel between the two remote membranes). However, for a tiny
membrane corresponding to $p_-=1/R$ (say, at the lowest level of the
DLCQ) there are no degrees of freedom to store the information on the
direction of the membrane, hence the $p_-=1/R$ describes the wave
function of the tiny membrane after averaging over all anti-self-dual
orientations. Hence the average signed volumes for the $p_-=1/R$
particles should vanish. In particular it should vanish for two remote
membranes at the $p_-=2/R$ level, which is not what we find. We would
like to suggest, however, that this is an artifact of the fact that we
are working at small $p_-$. Clearly the lack of ability to describe
independent orientations of the two remote membranes has to do with
the fact that the manifold is of very low dimension, i.e., low $p_-$.

\subsec{A Large Round Disk Solution}

We are going to discuss the simplest coplanar arbitrary $N$ solution
which corresponds to an open membrane whose image in spacetime
looks like a planar disk. The solution is

\eqn\dsksol{
X=\left(\matrix{ 0& \sqrt{N-1} \cr
		  &0&\sqrt{N-2} \cr
		  &&.		\cr
		  &&&0&1\cr
		  &&&& 0 }\right), \ \ \ 
		Q\propto \left(\matrix{ 0\cr 0\cr .\cr 1}\right),
\ \ {\tilde Q}={\tilde X}=0 }
This configuration has a $U(1)$ symmetry which mixes $X$ and
$X^\dagger$, suggesting that this membrane is the round disk. We
would like to compute the size of the membrane in the large N limit
and we will do so by computing its moments $<r^l>$ which are, to
leading order in $N$ (and neglecting overall coefficients)
\eqn\momclc{
Tr\bigl((1-P)[X,X^\dagger](1-P)(XX^\dagger)^l\bigr)\sim {1\over l+1}
N^{l+1}\ \ \ (at\ large\ N)} which indicates a uniform disk of radius
$R\propto \sqrt{N}$ (ordering ambiguities in the translation from
$r^2$ to $XX^\dagger$ are subleading in $N$).

This is also a good test case to examine our procedure for inserting
the projection operator $1-P$. Without inserting it, the results for
the moments would be
$$-N^{l+1}+{1\over l+1}N^{l+1}$$ which corresponds to a disk of
membrane with a ring of anti-membrane charge along its
perimeter. Again is a configuration which is difficult to interpret,
whereas the insertion of the projection operators yields precisely the
expected results.

\newsec{Discussion}

The D0-D4 system deformed by a FI term is the discrete light cone
quantization of the 6-dimensional (2,0) theory with a large $H_3$
field turned on (in a specific way), i.e., a theory which generalizes
non-commutative geometry to 3-forms. We showed how to naturally
interpret this model as configurations of open fluctuating membranes,
how to evaluate the membrane ground state wave function and suggested
how to measure some aspects of membrane size and shape.  The key to
this identification is the addition of the matter fields in the
fundamental representation of the $U(N)$ gauge symmetry - the
reparametrization of the sphere - which correspond to point like
impurities on the surface on which the surface can rip open and become
an open membrane. 

\bigskip

\centerline{\bf Acknowledgment}

I would like to thank O. Aharony, O. Ganor, A. Kapustin, M. Rozali,
N. Seiberg, M. Spradlin, A. Strominger, H. Verlinde and A. Volovich for useful
discussion. This work is supported by NSF grant 98-02484.
 
\listrefs

\end